\documentclass[a4paper]{report}
\usepackage[utf8]{inputenc}
\usepackage[T1]{fontenc}
\usepackage{RJournal}
\usepackage{amsmath,amssymb,array}
\usepackage{booktabs}
\usepackage{acronym}
\usepackage{thumbpdf}

\usepackage{Sweave}
\begin{document}

\sectionhead{Contributed research article}
\volume{XX}
\volnumber{YY}
\year{20ZZ}
\month{AAAA}

\begin{article}

\title{\pkg{tsmp}: An R Package for Time Series with Matrix Profile}
\author{by Francisco Bischoff and Pedro Pereira Rodrigues}

\maketitle

\abstract{
This article describes \CRANpkg{tsmp}, an \proglang{R} package that implements the \acl{MP} concept for \acl{TS}. The \CRANpkg{tsmp} package is a
toolkit that allows all-pairs similarity joins, motif, discords and chains discovery, semantic segmentation, etc. Here we describe how the \CRANpkg{tsmp} package may be used by showing some of the use-cases from the original articles and evaluate the algorithm speed in the \proglang{R} environment. This package can be downloaded at \url{https://CRAN.R-project.org/package=tsmp}.
}

\section{Introduction: time series data mining} \label{sec:intro}

A \ac{TS} is a sequence of real-valued numbers indexed in time order. Usually, this sequence is taken in a regular period of time, which will be assumed to be true in this context. The interests in \ac{TS} data mining have been growing along with the increase in available computational power. This kind of data is easily obtained from sensors (e.g., \ac{ECG}), (ir)regular registered data (e.g., weekly sales, stock prices, brachial blood pressure). Even other kinds of data can be converted to TS format, such as shapes \citep{Wei2006} and DNA sequences \citep{Shieh2008}. \ac{TS} are generally large, high dimensional and continuously updated which requires algorithms fast enough in order to be meaningful. Besides, unlike other kinds of data, which usually have exact answers, \ac{TS} are usually analysed in an approximated fashion.

These characteristics have been challenging researchers to find faster and more accurate methods to retrieve meaningful information from \ac{TS}. This required one or more of these methods: dimensionality reduction, constraints, domain knowledge, parameter tweaks. Only afterwards could the data mining tasks be applied in feasable time. Typical tasks include motif and discord discovery, subsequence matching, semantic segmentation, rule discovery, similarity search, anomaly detection, clustering, classification, indexing, etc. \citep{Fu2011}.

This paper describes the \CRANpkg{tsmp} package \citep{tsmp} which uses a novel approach to \ac{TS} data mining: the \ac{MP} \cite{Yeh2017a}, which is based on the \ac{APSS} (also known as similarity join). The \ac{APSS}' task is to, given a collection of data objects, retrieve the nearest neighbour for each object. The remaining part of this paper is organised as follows: In Section~\ref{sec:matrixprofile} we describe the reasoning behind the \ac{MP}, in Section~\ref{sec:thepkg} we present the \CRANpkg{tsmp} package with examples, in Section~\ref{sec:speed} we compare the performance of the \proglang{R} implementation, and in Section~\ref{sec:conclusion} we conclude with a brief discussion.

\section{The matrix profile} \label{sec:matrixprofile}

The reader may be aware of what a \ac{DM} is. It is widely used in \ac{TS} for clustering, classification, motif search, etc. But, even for modestly sized datasets, the algorithms can take months to compute even with speed-up techniques such as indexing \citep{Shieh2008,Fu2008}, lower-bounding \citep{Keogh2005}, data discretization \citep{Lin2003} and early abandoning \citep{Faloutsos1994}. At best, they can be one or two orders of magnitude faster.

The \ac{MP} is an ordered vector that stores the Euclidean distance between each pair within a similarity join set. One (inefficient) way would be to use the full \ac{DM} of every iteration of a sliding window join and retrieve just the smallest (non-diagonal) value of each row. The \ac{MP} also has a companion vector called \ac{PI}, that gives us the position of the nearest neighbour of each subsequence.

This method has a host of interesting and exploitable properties. For example, the highest point on the \ac{MP} corresponds to the \ac{TS} discord, the (tied) lowest points correspond to the locations of the best \ac{TS} motif pair, and the variance can be seen as a measure of the \ac{TS} complexity. Moreover, the histogram of the values in the \ac{MP} is the exact answer to the \ac{TS} density estimation. Particularly, it has implications for \ac{TS} motif discovery, \ac{TS} joins, shapelet discovery (classification), density estimation, semantic segmentation, visualisation, rule discovery, clustering, etc. \citep{Yeh2017a}.

Some of the advantages/features of this method:

\begin{itemize}
  \item It is \emph{exact}, providing no false positives or false dismissals.
  \item It is \emph{simple} and parameter-free. In contrast, the more general metric space \ac{APSS} algorithms require building and tuning spatial access methods and/or hash functions.
  \item It requires an inconsequential space overhead, just $ O(n) $ with a small constant factor.
  \item It is extremely \emph{scalable}, and for \emph{massive} datasets, we can compute the results in an anytime fashion, allowing ultra-fast \emph{approximate} solutions.
  \item Having computed the similarity join for a dataset, we can incrementally update it very efficiently. In many domains, this means we can effectively maintain exact joins on \emph{streaming} data forever.
  \item It provides \emph{full joins}, eliminating the need to specify a similarity threshold, which is a near-impossible task in this domain.
  \item It is \emph{parallelizable}, both on multicore processors and in distributed systems \citep{Zhu2016}.
\end{itemize}

\section[The tsmp package]{The \CRANpkg{tsmp} package} \label{sec:thepkg}

The \CRANpkg{tsmp} package provides several functions that allow for an easy workflow using the \acf{MP} concept for \ac{TS} mining. The package is available from the \ac{CRAN} at \url{https://CRAN.R-project.org/package=tsmp}. In Section~\ref{sec:instalation} we explain how to install this package. In Section~\ref{sec:input} we describe the syntax for the main functions in \CRANpkg{tsmp}, giving an example of a particular model. In Section~\ref{sec:computational} we will further explain the available algorithms for \ac{MP} computation and its current use. In Section~\ref{sec:datamining} we show some examples of \ac{MP} application for data mining.

\subsection{Installation} \label{sec:instalation}

The \CRANpkg{tsmp} package can be installed in two ways:

The release version from \ac{CRAN}:
\begin{Verbatim}
install.packages("tsmp")
\end{Verbatim}

or the development version from GitHub:
\begin{Verbatim}
# install.packages("devtools")
devtools::install_github("franzbischoff/tsmp")
\end{Verbatim}

\subsection{Input arguments and example} \label{sec:input}

The \CRANpkg{tsmp} has a simple and intuitive workflow. First, you must compute the \ac{MP} of the desired \ac{TS}. Depending on the task, the user might want to follow one of three paths: univariate self-join, AB-join or multivariate self-join. One exception is the SiMPle algorithm that is a multivariate AB-join and will be explained in Section~\ref{sec:computational}.

The main function is \code{tsmp()}, which has the following usage:
\begin{Verbatim}
tsmp(..., window_size, exclusion_zone = 1/2,
  mode = c("stomp", "stamp", "simple", "mstomp", "scrimp"),
  verbose = 2, s_size = Inf, must_dim = NULL, exc_dim = NULL,
  n_workers = 1, .keep_data = TRUE)
\end{Verbatim}

The first argument \code{ellipsis} (the three dots) receives one or two \ac{TS}. For self-joins, the user must input just one \ac{TS}; two for AB-joins. Multivariate \ac{TS} may be input as a matrix where each column represents one dimension. Alternatively, the user may input the Multivariate \ac{TS} as a list of vectors. The second argument \code{window\_size} is the size of the sliding window. These are the most basic parameters you need to set.

Further parameters are:

\begin{itemize}
  \item \code{exclusion\_zone}, is an important parameter for self-joins. This is used to avoid trivial matches and is a modifier of the \code {window\_size}, i.e., for an \code{exclusion\_zone} of $ 1/2 $, and \code{window\_size} of 50, internally the result will be 25.
  \item \code{mode}, here the user may choose the algorithm used for the \ac{MP} calculation. \code{stomp}, \code{stamp} and \code{scrimp} return equal results, although differing in some practical attributes, and they will be further explained in Section~\ref{sec:computational}. \code{mstomp} is designed for Multivariate \ac{TS} self-join only. \code{simple} is designed for Multivariate \ac{TS} for self-join and AB-join, which will also be further explained in Section~\ref{sec:computational}.
  \item \code{verbose}, controls the verbosity of the function. 0 means no feedback, 1 means text messages only, 2 (the default) means text messages and progress bar, and 3 also plays a sound when finished.
  \item \code{s\_size}, controls the \emph{anytime} algorithms. This is just a way to end the algorithm in a controlled manner because the \emph{anytime} algorithms can be stopped \emph{anytime} and the result will be returned.
  \item \code{must\_dim}, is an optional argument for the \code{mstomp} algorithm. See next item.
  \item \code{exc\_dim}, as \code{must\_dim}, is an optional argument for the \code{mstomp} algorithm. These arguments control which dimensions must be included and which must be excluded from the multidimensional \ac{MP}.
  \item \code{n\_workers}, controls how many threads will be used for the \code{stamp}, \code{stomp}, and \code{mstomp}. Note that for small datasets, multiple threads add an overhead that makes it slower than just one thread.
  \item \code{.keep\_data}, \code{TRUE} by default, keeps the input data inside the output object. This is useful for chained commands.
\end{itemize}

\subsubsection{Example data} \label{sec:example}

We think that the best and simple example to demonstrate the \CRANpkg{tsmp} package is the motif search.

The \CRANpkg{tsmp} package imports the \code{\%>\%} (pipe) operator from the \pkg{magrittr} package that makes the \CRANpkg{tsmp} workflow easier.

The following code snippet shows an example of the workflow for motif search:

\begin{Schunk}
\begin{Sinput}
R> data <- mp_fluss_data$walkjogrun$data
R> motifs <- tsmp(data, window_size = 80, exclusion_zone = 1/2) %>%
+    find_motif(n_motifs = 3, radius = 10, exclusion_zone = 20) %T>% plot()
\end{Sinput}
\end{Schunk}

The \code{find\_motif()} function is an S3 class that can receive as the first argument the output of \code{tsmp()} function as a univariate or multivariate \ac{MP}. This allows us to use the pipe operator easily. The \code{plot()} function is also an S3 class extension for plotting objects from the \CRANpkg{tsmp} package and works seamlessly.

\subsection{Computational methods} \label{sec:computational}

There are several methods to compute the \acf{MP}. The reason for that is the unquenchable need for speed of the UCR's researchers. Before starting, let's clarify that the time complexity of a brute force algorithm has a time complexity of $ O(n^{2}m) $, for $ n $ being the length of the reference \ac{TS} and $ m $ the length of the sliding window (query) that is domain dependent.

\subsubsection[STAMP]{Scalable time series anytime matrix profile (\acsu{STAMP})} \label{sec:stamp}

This was the first algorithm used to compute the \ac{MP}. It uses the \ac{MASS} \citep{Mueen2015} as the core algorithm for calculating the similarity between the query and the reference \ac{TS}, called the \ac{DP}. The ultimate \ac{MP} comes from merging the element-wise minimum from all possible \acp{DP}. This algorithm has the time complexity of $ O(n^{2}\log n) $ and space complexity of $ O(n) $ \citep{Yeh2017a}. The \emph{anytime} property is achieved using a random approach where the best-so-far \ac{MP} is computed using the \acp{DP} that have been already calculated.

\subsubsection[STOMP]{Scalable time series ordered-search matrix profile (\acsu{STOMP})} \label{sec:stomp}

This was the second algorithm used to compute the \ac{MP}. It also uses the \ac{MASS} to calculate the \ac{DP} but only for the first iteration of each batch. The researchers noticed that they could reuse the values calculated of the first \ac{DP} to make a faster calculation in the next iterations. This results on a time complexity of $ O(n^{2}) $, keeping the same space complexity of $ O(n) $. This algorithm is also suitable for a \ac{GPU} framework (although this was not yet implemented in \CRANpkg{tsmp} package) \citep{Zhu2016}. The main drawback of \ac{STOMP} compared with \ac{STAMP} is the lack of the \emph{anytime} property. In scenarios where a fast convergence is needed (e.g., finding the top-$k$ motifs) it may be required only 5\% of the \ac{MP} computation to provide a very accurate approximation of the final result.

\subsubsection[SCRIMP]{Scalable column independent matrix profile (\acsu{SCRIMP})} \label{sec:scrimp}

The \ac{SCRIMP} algorithm is still experimental at the time of this article. It combines the best features of \ac{STOMP} and \ac{STAMP}, having a time complexity of $ O(n^{2}) $ and the \emph{anytime} property \citep{Ucrmp}.

\subsubsection[SiMPle]{Similarity matrix profile (\acsu{SiMPle})} \label{sec:simple}

The \ac{SiMPle} algorithm is a variation designed for music analysis and exploration \citep{Silva2018}. Internally it uses \ac{STOMP} for \ac{MP} computation and allows multidimensional self-joins and AB-joins. The resulting \ac{MP} is computed using all dimensions. One major difference is that it doesn't apply any z-normalization on the data, since for music domain this would result in spurious similarities.

\subsubsection[mSTOMP]{Multivariate \acsu{STOMP} (\acsu{mSTOMP})} \label{sec:mstomp}

The \ac{mSTOMP} algorithm was designed to motif search in multidimensional data \citep{Yeh}. Performing motif search on \emph{all} dimensions is almost guaranteed to produce meaningless results, so this algorithm, differently from \ac{SiMPle}, doesn't compute the \ac{MP} using all dimensions na\"ively, but the $ d $-dimensional \acp{MP} for every possible setting of $ d $, simultaneously, in $ O(dn^{2} \log d) $ time and $ O(dn) $ space. The resulting \acp{MP} allow motif search in multiple dimensions and also to identify which dimensions are relevant for the motifs founded.

\subsection{Data mining tasks} \label{sec:datamining}

\subsubsection{Motif search} \label{sec:motifsearch}

In Section~\ref{sec:example} we have shown a basic example of the workflow for motif search. Let's take a look at the result of that code:

\begin{Schunk}
\begin{Sinput}
R> motifs
\end{Sinput}
\begin{Soutput}
Matrix Profile
--------------
Profile size = 9922 
Window size = 80 
Exclusion zone = 40 
Contains 1 set of data with 10001 observations and 1 dimension 

Motif
-----
Motif pairs founded = 3 
Motif pairs indexes = [584, 741] [3917, 5088] [8543, 8843]  
Motif pairs neighbors = [2948] [6719] [6849, 4944]  
\end{Soutput}
\end{Schunk}

As we can see, this is a summary that \CRANpkg{tsmp} package automatically generates from the resulting object. One nice property is that the object always holds the original \ac{MP} and by default also holds the input data so that you can keep mining information from it. If the dataset is too big or you are concerned about privacy, you may set the argument \code{.keep\_data = FALSE}.

In addition to this summary, you can see the results using \code{plot()} in Figure~\ref{fig:motif}:

\begin{figure}[!htb]
\centering
\begin{Schunk}
\begin{Sinput}
R> plot(motifs, type = "matrix")
\end{Sinput}
\end{Schunk}
\includegraphics{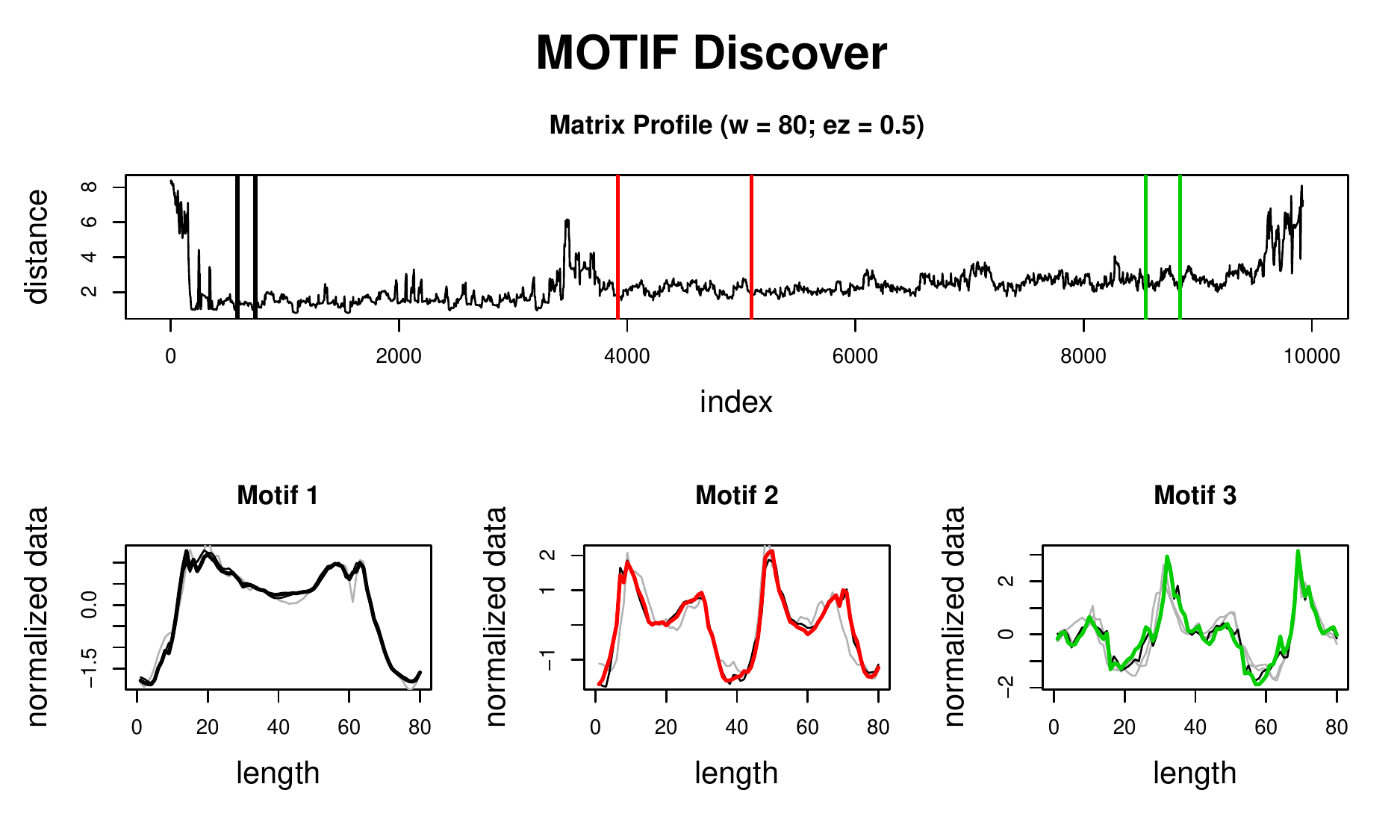}
\caption{\label{fig:motif} The upper graphic shows the computed \ac{MP} with each motif pair as a coloured vertical bar. The lower graphics show each motif in colour and the founded neighbours in grey.}
\end{figure}

\newpage

This dataset is the \emph{WalkJogRun} PAMAP's dataset \citep{Reiss2012}. It contains the recording of human movements in three states, walking, jogging and running. As we can see, the plot shows the motifs of each state. Experienced readers might say that this is not the purpose of motif search, and we agree. The result shown here was achieved using a large \code{radius} and \code{exclusion\_zone} to force the algorithm to look for distant motifs. Semantic segmentation is the proper algorithm for this task, and we will show this in the next section.

\subsubsection{Semantic segmentation} \label{sec:semseg}

As previously explained, the resulting object holds the original data and \ac{MP}. So let's save some time and use the resulting object from the last section to try to find where the human subject started to jog and to run:

\begin{Schunk}
\begin{Sinput}
R> segments <- motifs %>% fluss(num_segments = 2)
R> segments
\end{Sinput}
\begin{Soutput}
Matrix Profile
--------------
Profile size = 9922 
Window size = 80 
Exclusion zone = 40 
Contains 1 set of data with 10001 observations and 1 dimension 

Arc Count
---------
Profile size = 9922 
Minimum normalized count = 0.063 at index 3448 

Fluss
-----
Segments = 2 
Segmentation indexes = 3448 6687 
\end{Soutput}
\end{Schunk}

\begin{figure}[!htb]
\centering
\begin{Schunk}
\begin{Sinput}
R> plot(segments, type = "data")
\end{Sinput}
\end{Schunk}
\includegraphics{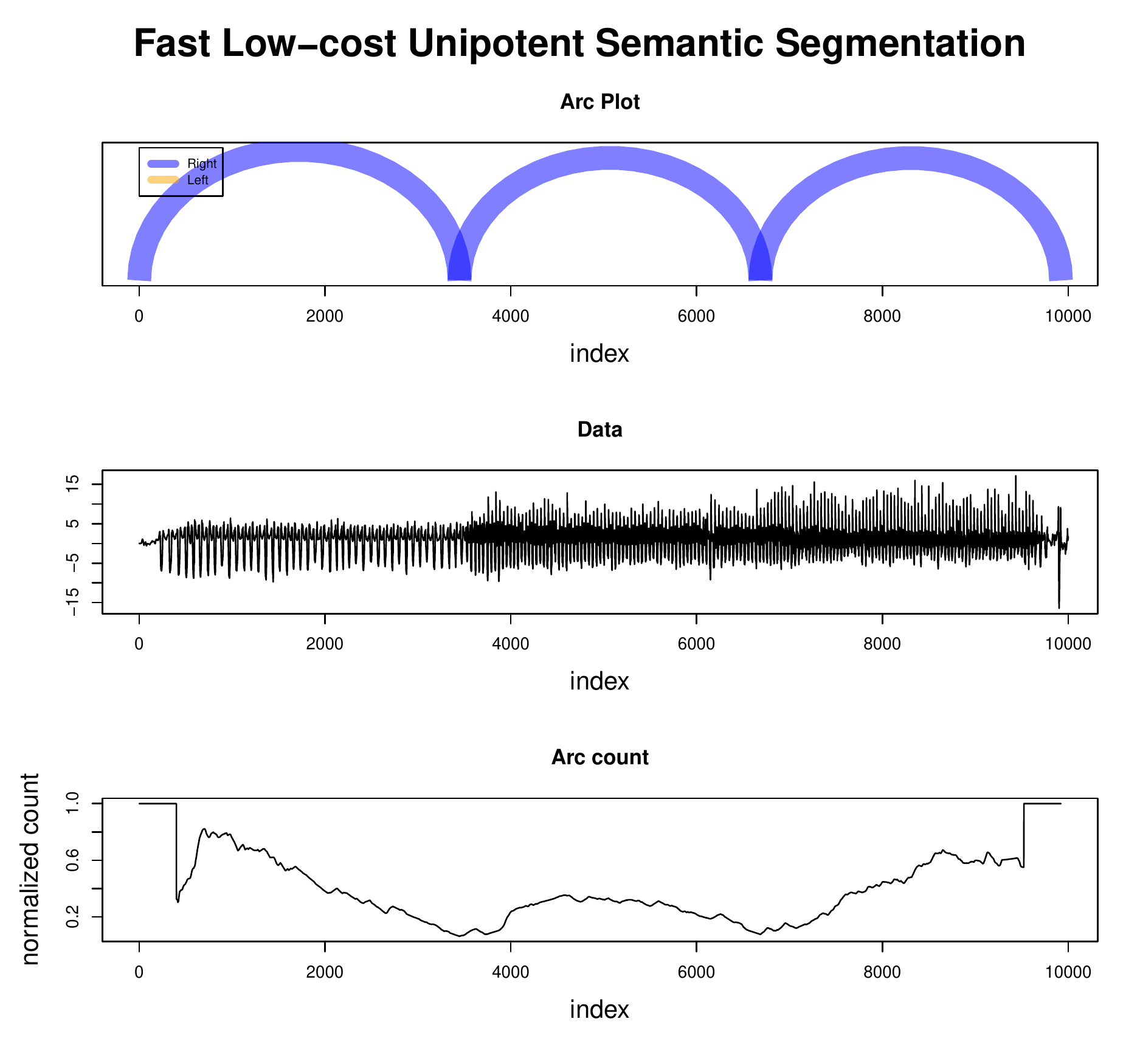}
\caption{\label{fig:segments} Semantic segmentation using \ac{MP}. The upper graphic shows the arc plot of predicted semantic changes (ground truth is 3800 and 6800). The middle graphic shows the data. The lower graphic shows the normalised arc counts with correction for the "edge-effect" \citep{Gharghabi2017}.}
\end{figure}

We can see that this object now holds information of the \ac{FLUSS} algorithm \citep{Gharghabi2017}, but the motif information is still there and can be retrieved using \code{as.motif()}. In Figure~\ref{fig:segments} we can see the graphic result of the segmentation.

\newpage

\subsubsection{Time series chains} \label{sec:chains}

As a final example of practical application, let's search for a new kind of primitive: time series chains \citep{Zhu2018}. This algorithm looks for patterns that are not just similar but evolve through time. The dataset used in this example is a record of the Y-axis of a mobile phone accelerometer while placing it on a walking subject's pocket \citep{Hoang2015}. The authors of this dataset wanted to analyse the stability of the mobile phone as it slowly settles in the pocket. This is a good example of a pattern that changes through time. Let's start with the workflow for this example:

\begin{Schunk}
\begin{Sinput}
R> chains <- mp_gait_data %>% tsmp(window_size = 50, exclusion_zone = 1/4,
+    verbose = 0) %>% find_chains()
R> chains
\end{Sinput}
\begin{Soutput}
Matrix Profile
--------------
Profile size = 855 
Window size = 50 
Exclusion zone = 13 
Contains 1 set of data with 904 observations and 1 dimension 

Chain
-----
Chains founded = 58 
Best Chain size = 6 
Best Chain indexes = 148 380 614 746 778 811 
\end{Soutput}
\end{Schunk}

\begin{figure}[!htb]
\centering
\begin{Schunk}
\begin{Sinput}
R> plot(chains, ylab = "")
\end{Sinput}
\end{Schunk}
\includegraphics{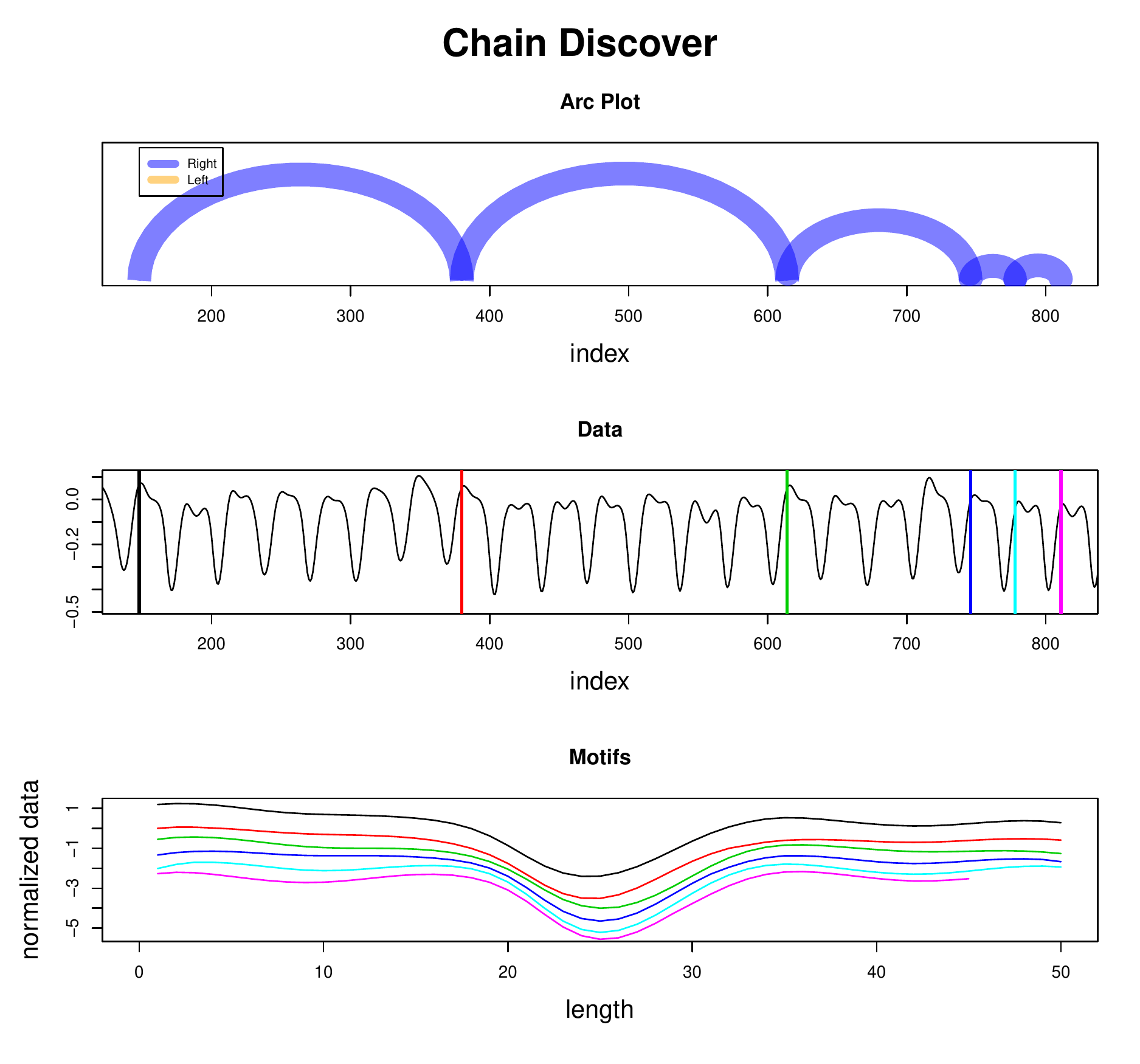}
\caption{\label{fig:chains} Finding evolving patterns using \ac{MP}. The upper graphic shows the arc plot of the discovered patterns. The middle graphic shows the data and the position of every pattern as a vertical coloured line. The lower graphic shows the patterns for comparison. They are y-shifted for visualisation only.}
\end{figure}

Here we see that the algorithm found 58 chains. \emph{Id est}, it found 58 evolving patterns with at least three elements, and the best one is presented in the last line, a chain with six elements. Figure~\ref{fig:chains} shows the patterns discovered.

\section{Speed} \label{sec:speed}

While this new method for \ac{TS} data mining is extremelly fast, we have to take into consideration that the \proglang{R} environment is not as fast as a low-level implementation such as \proglang{C/C++}. In Table~\ref{tab:speedr} we present the comparison to the \proglang{MATLAB} version that is available at the \ac{UCR}. \citet{Yeh2017a} shows that the slowest algorithm (STAMP) can be hundreds of times faster than the MK algorithm (the fastest known \emph{exact} algorithm for computing \ac{TS} motifs) \citep{Yoon2015}, while the \proglang{R} implementation is just 1.65 to 8.04 times slower than \proglang{MATLAB}'s, which is not a problem for an \proglang{R} researcher.

\begin{Schunk}
\begin{Sinput}
R> set.seed(2018)
R> data <- cumsum(sample(c(-1, 1), 40000, TRUE))
\end{Sinput}
\end{Schunk}

\begin{table}[!htb]
\centering
\begin{tabular}{cccc}
\hline
Algorithm & \proglang{R} Time$^*$ & \proglang{MATLAB} Time$^*$ & Threads \\ \hline
scrimp    &     \hphantom{0}45.30 &          \hphantom{0}27.49 &       1 \\
stomp     &     \hphantom{0}52.72 &          \hphantom{0}10.27 &       8 \\
stomp     &                136.01 &          \hphantom{0}16.91 &       1 \\
stamp     &                140.25 &          \hphantom{0}55.57 &       8 \\
stamp     &                262.03 &                     113.18 &       1 \\ \hline

\end{tabular}
\caption{\label{tab:speedr} Performances of \proglang{R} and \proglang{MATLAB} implementations on an Intel(R) Core(TM) i7-7700 CPU @ 3.60GHz using a random walk dataset. $^*$Median of 5 trials, in seconds.}
\end{table}

\section{Conclusion} \label{sec:conclusion}

The examples in Section~\ref{sec:datamining} show how straightforward the usage of \CRANpkg{tsmp} package is. Regardless, these examples are just a glimpse of the potential of the \ac{MP}. Several new algorithms based on \ac{MP} are being developed and will be gradually implemented in the \CRANpkg{tsmp} package \citep{Linardi2018,Zhu2018a,Gharghabi2018,Imani2018}. \cite{Yeh} for example, have developed an algorithm to allow \ac{MDS} visualisation of motifs. \cite{Gharghabi2018} have developed a new distance measure that better suits repetitive patterns \citep{Imani2018}.

The \ac{MP} has the potential to revolutionise the \ac{TS} data mining due to its generality, versatility, simplicity and scalability \citep{Ucrmp}. All existing algorithms for \ac{MP} have been proven to be flexible to be used in several domains using very few parameters and they are also robust, showing good performance with dimensionality reduced data and noisy data. In addition, a yet to be published article shows a fantastic score of $ >10^{18} $ pairwise comparisons a day using GPU for motif discovery \citep{Zimmerman}.

The \CRANpkg{tsmp} package is the first known \ac{MP} toolkit available on any statistical language, and we hope it can help researchers to better mining \ac{TS} and also to develop new methods based on \ac{MP}.


\section{Acknowledgements}

We would like to thank the researchers from \ac{UCR} for their contribution and permission to use their base code to be implemented in this package. Particularly to Prof. Eamonn Keogh whose work and assistance led to this project. We also acknowledge the participation in project NanoSTIMA (NORTE-01-0145-FEDER-000016) which was financed by the North Portugal Regional Operational Program (NORTE 2020) under the PORTUGAL 2020 Partnership Agreement and through the European Regional Development Fund (ERDF).

\newpage

\section{Acronyms}

\begin{acronym}[NNNNNN] 
  \acro{APSS}{all-pairs similarity search}
  \acro{CRAN}{Comprehensive \proglang{R} Archive Network}
  \acro{DM}{distance matrix}
  \acro{DP}{distance profile}
  \acro{ECG}{electrocardiogram}
  \acro{FLUSS}{fast low-cost unipotent semantic segmentation}
  \acro{GPU}{graphics processor unit}
  \acro{MASS}{Mueen's algorithm for similarity search}
  \acro{MDS}{multidimensional space}
  \acro{MP}{matrix profile}
  \acro{mSTOMP}{multivariate scalable time series ordered-search matrix profile}
  \acro{PI}{profile index}
  \acro{SCRIMP}{scalable column independent matrix profile}
  \acro{SiMPle}{similarity matrix profile}
  \acro{STAMP}{scalable time series anytime matrix profile}
  \acro{STOMP}{scalable time series ordered-search matrix profile}
  \acro{TS}{time series}
  \acro{UCR}{University of California Riverside}
\acresetall

\end{acronym}

\newpage

\bibliography{refs}

\address{Francisco Bischoff\\
  CINTESIS - Center for Health Technology and Services Research\\
  MEDCIDS - Community Medicine, Information and Health Decision Sciences Department\\
  Faculty of Medicine of the University of Porto\\
  Rua Dr. Placido Costa, s/n\\
  4200-450 Porto, Portugal\\
  ORCiD: 0000-0002-5301-8672\\
  \email{fbischoff@med.up.pt}}

\address{Pedro Pereira Rodrigues\\
  CINTESIS - Center for Health Technology and Services Research\\
  MEDCIDS - Community Medicine, Information and Health Decision Sciences Department\\
  Faculty of Medicine of the University of Porto\\
  Rua Dr. Placido Costa, s/n\\
  4200-450 Porto, Portugal\\
  ORCiD: 0000-0001-7867-6682\\
  \email{pprodrigues@med.up.pt}}

\end{article}

\end{document}